\documentstyle[twocolumn,twoside,epsf,rotate]{article}
\newcommand{\hb}{\\ \hspace*{2ex}}

\def\Mov{M_{\mathrm{OV}}}
\def\yr#1 {\relax}
\def\v#1 {{\bf#1.}, }
\def\p#1 {#1.}
\def\t#1 {{\bf#1.}, }
\def\s#1 {#1.}

\def\AZh{{\it Astron. Zhurn.}, }
\def\ApJ{{\it Ap.J.}, }
\def\AA{{\it As.Ap.}, }
\def\UFN{{\it Phys.Usp.}, }
\def\PAZh{{\it Pisma Astron. Zhurn.}, }
\def\Nat{{\it Nature}, }
\def\MNRAS{{\it MNRAS}, }

\begin{document}
\title{WHY NS AND BH MASS DISTRIBUTION IS BIMODAL?}

\author{M.E.\,Prokhorov, K.A.\,Postnov\\[2mm]
 Sternberg Astronomical Institute, Moscow State University,\hb
 Universitetskii pr. 13, Moscow 119899 Russia,\hb
{\em mystery@sai.msu.ru; pk@sai.msu.ru}\\[2mm]
}
\date{}
\maketitle

%

ABSTRACT.

The observed mass distribution for the compact remnants of massive
stars (neutron stars and black holes) and its relationship to
possible mechanisms for the ejection of the envelopes
of type II and Ib/c supernovae is analyzed. The conclusion is drawn that 
this distribution can be obtained only by a magneto-rotational
mechanism for the supernovae with sufficiently long time
of the field amplification, and a soft equation of state
for neutron stars with limiting masses 
$\sim$1.5--1.6~$M_\odot$.  Some consequences of this 
hypothesis are discussed. 
\\[1mm]
\\[2mm]

{\bf 1. Introduction}\\[1mm]

The observed masses of white dwarfs lie in a wide range from 
several tenths of a solar mass to nearly teh Chandrasekhar limit 
($\sim$1.2M$_\odot$), with low-mass white dwrafs encountered more often. 
We are not concerned with these objects here, and will not consider 
them further. The masses of neutron stars (NS) measured so far
lie within a very narrow interval: the masses for 26 NS radio pulsars
in binary systems are consistent with a normal distribution with mean mass  
1.35M$_\odot$ and dispersion 0.04M$_\odot$
(Thorsett and Chacrabarty 1999).
As noted by Thorsett and Chakrabarty, there is currently not a 
single pulsar in a binary system whose mass exceeds  
1.45M$_\odot$. The recently obtained upper limit
on the NS mass in  millisecond
pulsar PSR~J2019+2425 is $M_{\mathrm{NS}} < 1.51$~M$\odot$
(Nice et al 2001). If we add the less accurately determined
masses of NS in X-ray binaries  
(Cherepashchuk 2000) to this sample, the observed 
mean mass of NS is 
$M_{\mathrm{NS}} = (1.35\pm0.15)$~M$\odot$ (the same mean 
as above with a larger dispersion).

More than a dozen of black hole (BH) candidates in close
X-ray binary systems are known 
(see Cherepashchuk (1996, 2000) and references therein).
The masses of these objects are determined using 
radial velocity curves of optical counterparts of
binary systems. According to current data, the 
masses of BH candidates fill the interval  
$\sim$3--40~M$_\odot$, with a mean value of about
10~M$_\odot$. 

In addition to the reliable dynamical determinaitons of the NS and BH
masses in binary pulsars and X-ray novae, there are a number of less
accurate mass estimates for compact objects in X-ray binaries. 
(1) The mass of NS in the low-mass X-ray binary 
Cyg~X-2 is determined by Orosz and Kuulkers (1999) to be
$1.8\pm0.2$~M$_\odot$. (2) X-ray pulsar Vela~X-1: $M_{NS}\sim
1.9$~M$_\odot$, according to van~Kerkwijk et al (1995), but
$M_{NS}\sim 1.4$~M$_\odot$ according to Stickland et al (1997). (3) 
The eclipsing low-mass X-ray binary 
4U~1700--37:  
$M_{NS}=1.8\pm 0.4$~M$_\odot$
according to Heap and Corcoran (1992), 
but it could be a low-mass 
BH (Brown et al 1996). Until the high
masses of these NS are independently verified, 
we will consider them to be uncertain.

Thus, we assume that current reliable 
measurements of NS masses lie in a narrow
interval
$M_{\mathrm{NS}} = (1.35\pm0.15)$~M$_\odot$, 
masses of BH lie in a wide range 
M$_{\mathrm{BH}}>3$M$_\odot$, and {\it not a single} 
NS has currently been reliably detected to have mass
in the gap between 1.5 and 3M$_\odot$,
and the number of BH with such small masses is
small (the total absence of such BH {\it is not required}).

This picture is in a dramatic disagreement with both
a monotonic distribution of the initial masses of 
main-sequence stars and the monotonic distribution of
of masses of carbon and iron cores that are developed
during nuclear evolution of massive stars. 
If a massive star is deprived of its hydrogen 
envelope during evolution, its carbon
core is observed as a Wolf-Rayet star. The current
observations show that their masses lie in a wide
range from  
$\sim$3M$_\odot$ to $\sim$50M$_\odot$ (Cherepashchuk 2001, 2000, 1998). 
According to calculations
Timmes et al (1996), the masses of the iron cores before
collapse lie in the interval from 
1.25 to 2.05M$_\odot$. Both the carbon and iron cores
depend monotonically on the initial masses of the stars.
\\[2mm]

{\bf 2. The enevelope ejection}\\[1mm]

In sufficiently massive stars
($>$8--10M$_\odot$), which can produce NS and BH,
the nuclear evolution ends up with the core collapse
which can be accompanied by the envelope ejection, 
leading to the supernova type II or Ib/c.
If the shell is ejected "efficiently"
(i.e., it receives an energy of the order of
the bininding energy of the remnant),
it expands in the surrounding medium and a low-mass compact
object forms with a mass of order the mass of the collapsed
core of the pre-supernova. If the shell is ejected 
"inefficiently", a large fall of matter from the
envelope to the forming compact object is inevitable. 
As a result, the mass of the latter can substantially grow
and approach the pres-supernova mass.

There can be a continuous transition between these two limiting
cases. However, if we suppose that 
\emph{the ejection of the envelope during the supernova explosion 
is sharply \emph{(even in a step-like manner)}
weakened for pre-supermnova core masses above some threshold}, 
the continuous sequence of the pre-supernova masses 
would give rise to two types of objects with sufficiently
different masses.
\\[2mm]

{\bf 3. The core collapse}\\[1mm]

The formation of a compact object during the core collapse 
can occur in two ways. \\
(1) The direct collapse into a BH, bypassing an intermediate stage
of a hot proto-neutron star, if its mass  is above some threshold
$M_{dir}>\Mov$ (see Prakash et al (2000) and references therein
for a more detailed description of this process).\\
(2)
Via the intermediate stage with hot proto-neutron star
lasting several seconds or tens of seconds, in which there is 
intense radiation of thermal energy by the neutrino flux, 
after which the hot proto-neutron star "cools" or, if 
its mas exceeds the Oppenheimer-Volkoff limit for 
neutron star matter $\Mov$, collapses into a black hole. 

The modern calcupations of core collapses show that 
$M_{dir}-\Mov\approx0.3$--1M$_\odot$ 
(Strobel and Weigel (2000) and refrences therein).
Clearly, that for a  
\emph{static} NS $M_{dir}$ is always larger than
$\Mov$.

\begin{figure}[htbp]
\centerline{\epsfxsize=0.5\columnwidth\epsfbox{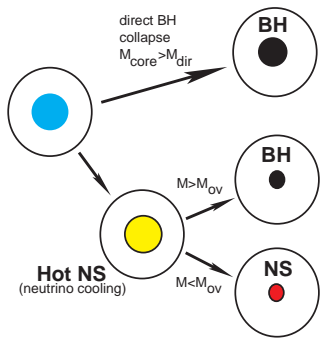}}
\caption{The scheme of the core collapse}
\label{f:collapse}
\end{figure}

The possible ways of the core collapse are schematically 
shown in Fig.~\ref{f:collapse}.
\\[2mm]

{\bf 4. Supernova mechanisms}\\[1mm]

Let us now consider various mechanisms for su-
pernova explosions to find a qualitative shape of the resulting mass 
distribution of compact objects. The consideration will be based
on the illustration with the scheme of Fig. 1 to the left, completed
with details of various SN mechanisms, and by the compact 
mass distribution plot turned counterclockwise to the right.  

Excluding exotic models, currently there are three different   
mechanisms for supernova explosions: (1) The standard mechanism, 
in which a shock wave appears as a result of the bounce of
the matter flux from the "solid" core; the shock wave propagation
is sustained by the neutrino flux. (2) The mechanism proposed by
Imshennik (1992) is associated with the division of the rapidly 
rotating collapsing stellar core into two parts. (3) Magneto-rotational
mechanism of envelope ejection (Bisnovatyi--Kogan 1970). 
Let us consider these mechanisms in turn. 

Note that none of these mechanisms can presently explain all 
the facts related to supernova phenomenon. So 
{\it a priori\/} all these mechanism may be 
equally applicable. 
\\[2mm]

{\it 4.1. The standard (neutrino) supernova mechanism}\\[1mm]

In the standard model, the energy is transferred from 
the hot compact remnant to the envelope by the neutrino flux. 
Unfortunately, this mechanism is unable to eject the supernova
shell either in the 
spherically symmetric or the axially symmetric (with rotation)
case (Janka 2001).
There is some hope that the situation can be saved
by large-scale neutrino convection  
(Herant et al 1994, Mezzacappa et al 1998). 

\begin{figure}[htbp]
\centerline{\epsfysize=1.0\columnwidth{\rotate[r]{\epsfbox{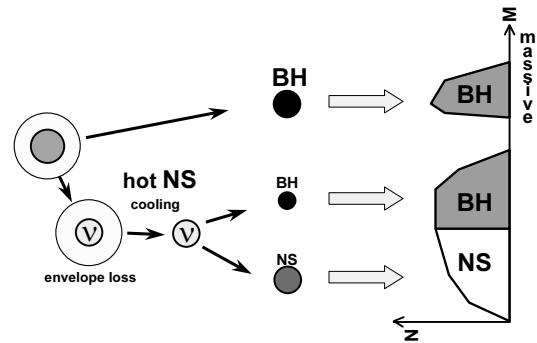}}}}
\caption{The standard (neutrino) mechanism}
\label{f:nu}
\end{figure}

The envelope ejection, if any, must occur on the first
stage of the hot NS with most intensive neutrino emission
(this stage lasts for several seconds). The neutrino 
fluxes from hot NS with a mass below and above 
$\Mov$ are not strongly different. In a
direct collapse, the hot stage is appreciably shorter (of
the order of the dynamical time scale for the collapse),
and, consequently, is less efficient.

The resulting mass distribution consists of a class of massive BH formed 
during the direct collapse and a comparable number of NS and 
low-mass BH (see Fig.~\ref{f:nu}). 
This distribution does not match with observations.
\\[2mm]

{\it 4.2. Imshennik's mechanism (the double core)}\\[1mm]

This mechanism is associated with the division of 
a rapidly rotating collpasing stellar core into two parts, 
at least one of which must be a NS. The parts of
the binary core then approach due to the emission
of gravitational radiation, until the component with
the smaller mass (and larger size) fills its Roche lobe.
Further, there is an exchange of mass until the mass
of the smaller component reaches the lower limit for
the mass of a neutron star (about 0.1$M_\odot$), at which
point there is an explosive de-neutronization of the
low-mass neutron star. This mechanism was first suggested by 
Blinnikov et al (1984) and applied to supernova explosions by 
Inshennik (1992). 
This additional release
of energy fairly far from the center of the collapsing
star can efficiently eject its envelope. This mechanism
can act only for the most rapidly rotating supernova
precursors.

\begin{figure}[htbp]
\centerline{\epsfysize=1.0\columnwidth{\rotate[r]{\epsfbox{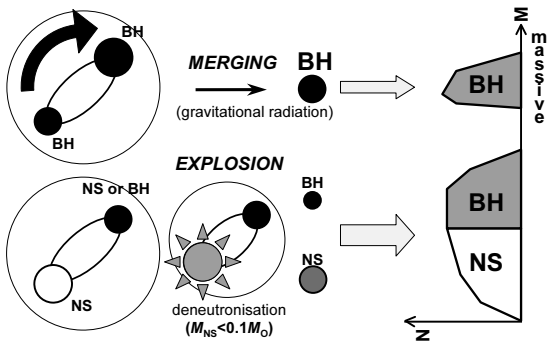}}}}
\caption{Imshennik's supernova mechanism}
\label{f:bin}
\end{figure}

The approach of the binary core up to its merging
could last from several minutes to several hours; 
i.e., appreciably longer than the hot neutron
star can exist. The scheme of the collapse shown in 
Fig.~\ref{f:bin} is somewhat different: 
here \emph{massive} is the binary system in which 
both parts of the core become BH. In this case 
the process results in a "quiet" coalescence of BH
with a concomitant accretion of matter from the envelope. 

The processes which take place in the 
\emph{low-mass} situation are descrivbed above
and their result is weakly dependent on the mass and type 
of the compact remnant. As a result, the same 
mass distribution as for the neutrino mechanism is obtained, 
in contradiction with observations.   
\\[2mm]

{\it 4.3. Magneto-rotational mechanism}\\[1mm]

This mechanism was proposed by Bisnovatyi-Kogan (1970).
The supernova shell is expelled by the magnetic field 
at the expense of the rotational energy of the newborn NS. The process
occurs in two stages. At the first stage, a toroidal magnetic field
appears and linearly grows with time. The duration of this
stage depends on the NS rotational velocity and its initial 
magnetic field value and can vary from fraction of a second to 
hours. When the magnetic field strength approaches some critical value 
($\sim10^{16}$--$10^{17}$~G), the magneto-rotational 
explosion occurs which accelerates and expells the 
envelope in 0.01--0.1~s 
(see Ardeljan et al (1998)). 
For this mechanism to operate, the star should have a
sufficiently rapid (but not limiting) rotation. 

Depending on the relation between the time of the magnetic field amplification 
$t_B$ (time before the explosion) and the hot NS cooling time scale
$t_\nu$, different compact object mass distributions appear. 

\begin{figure*}[htbp]
\centerline{%
\hbox to\columnwidth{\hss%
\epsfysize=1.0\columnwidth\rotate[r]{\epsfbox{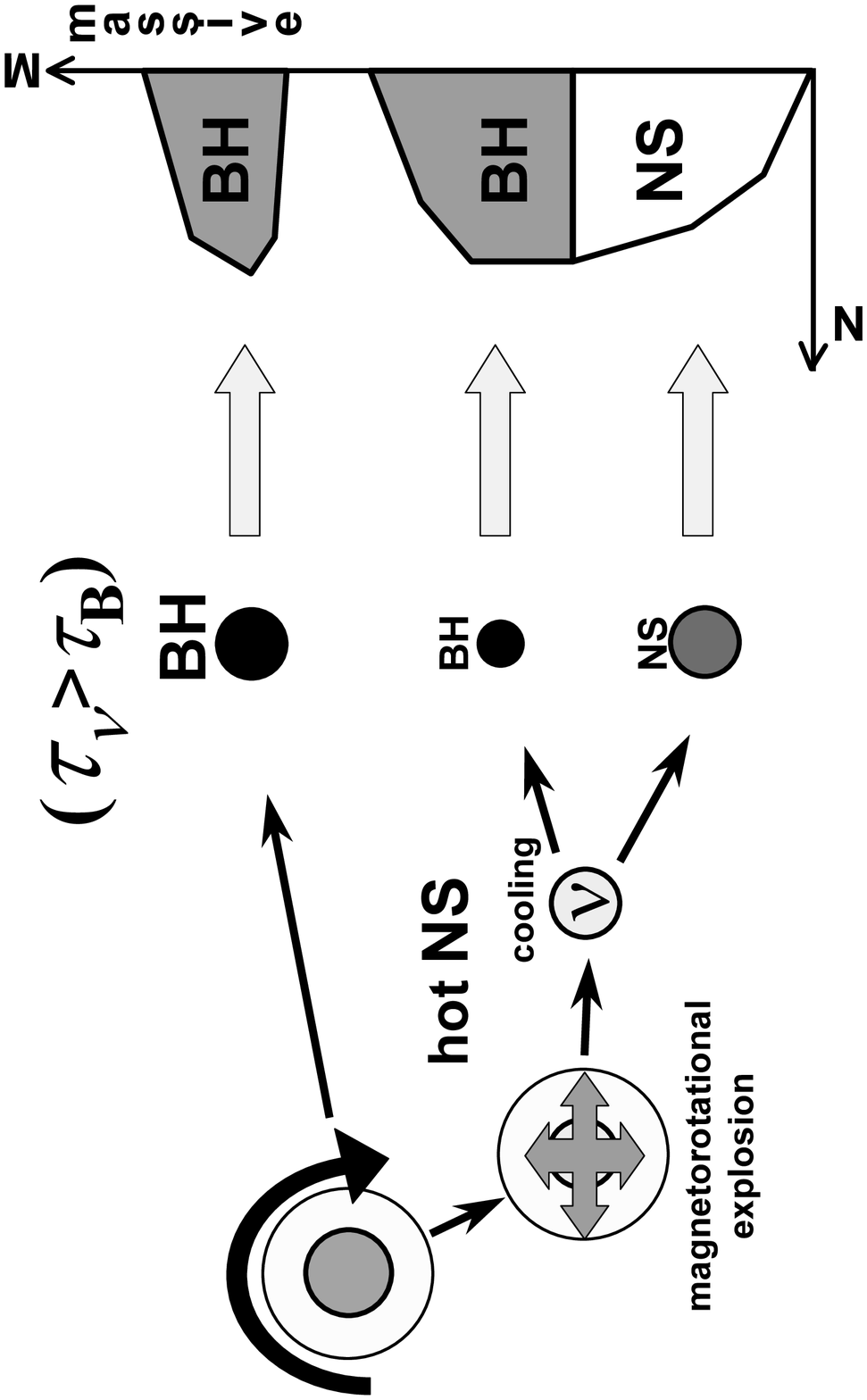}}
\hss}\hss%
\hbox to\columnwidth{\hss%
\epsfysize=1.0\columnwidth\rotate[r]{\epsfbox{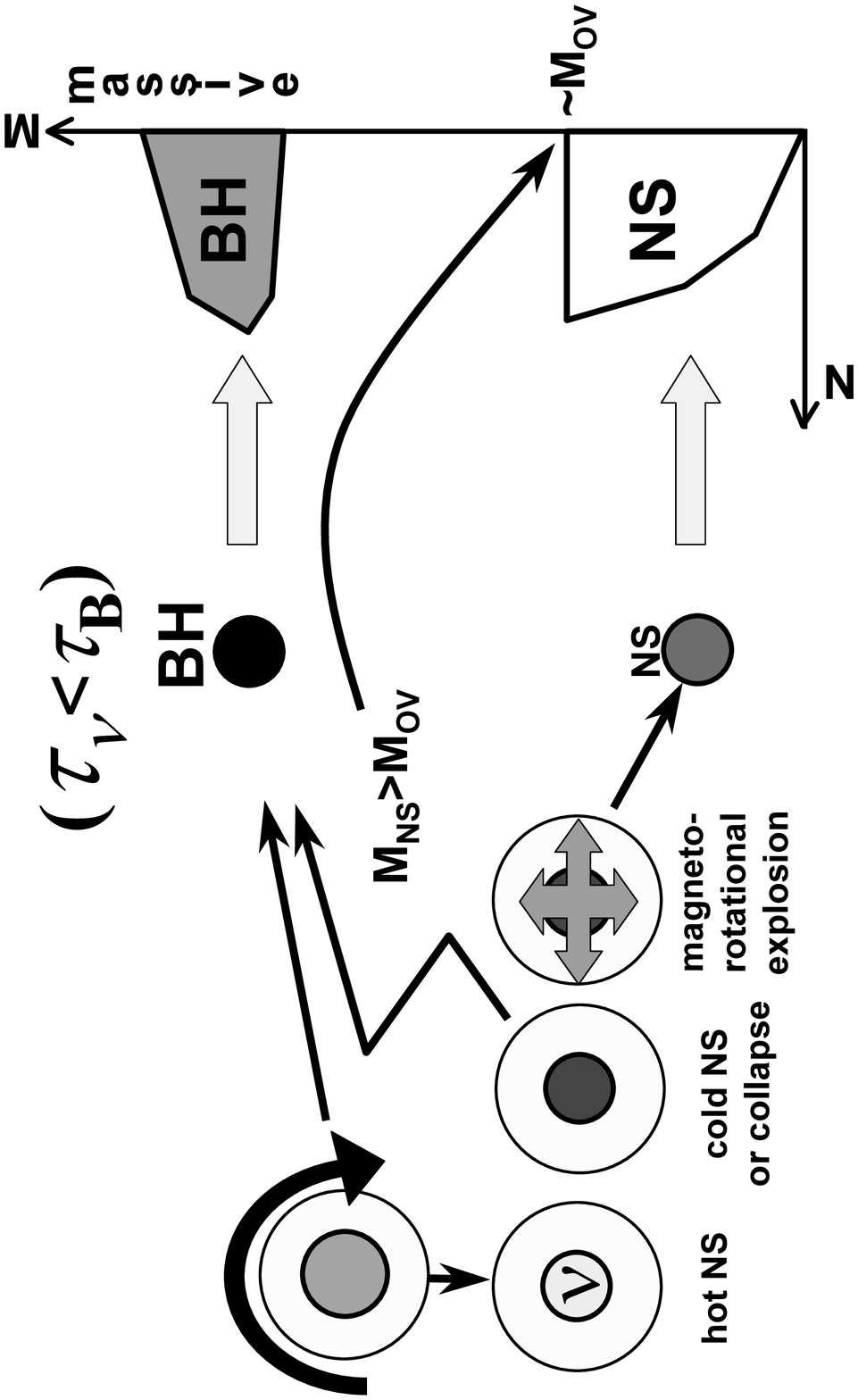}}
\hss}}
\caption{Two ways of the magneto-rotational mechanism for SN explosion}
\label{f:mr}
\end{figure*}

During the direct core collapse into BH, the magnetic field
amplification never starts and the envelope are not
ejected. In contrast, on the branch leading to NS formation,
the magneto-rotational mechanism ultimately leads to the explosion
and the envelope ejection.    

The difference between the two variants concerns only 
objects with masses
$\Mov < M < M_{dir}$, in which initially a hot NS forms and
after cooling collapses into BH. If the explosion occurs 
at the stage of a hot NS ($t_B < t_\nu$), the envelope 
is ejected before stars with 
$M>\Mov$ collapse; they form low-massive BH.
Therefore, we for the third time obtain the mass distribution 
in disagreement with observations. 

In contrast, if the field amplification porceeds slowly
($t_B > t_\nu$), the objects with 
$\Mov < M < M_{dir}$ collapse into BH, after which the
field amplification stops. No magneto-rotational explosion, 
and hence, envelope ejection 
occurs, so masses of these BH will be weakly different 
from those formed during the direct collapse. The mass
distribution will consist of only two groups of objects:
NS and massive BH.

Note one very important corollary of the scheme considered: the upper
boundary of the NS mass distribution must coincide (to the mass defect)
with $\Mov$. 
Thus the current observations suggest that NS should have  
\emph{a very soft} EOS with $\Mov\simeq1.5$--1.6$M_\odot$ (such 
equation are possible, e.g.
GS1, PAL6 and PCL2 in Lattimer and Prakash (2000).

The magneto-rotational explosion is illustrated in Fig.~\ref{f:mr}.
\\[2mm]

{\bf 5. Conclusions}\\[1mm]

We have shown that the magneto-rotational 
mechanism for supernova explosions, with the additional
requirements that the time before the explosion is
larger than the cooling time of a proto-NS
($\tau_B>\tau_\nu$) with a soft equation of state
of NS matter ($\Mov\simeq1.5$--$1.6~M_\odot$), 
leads naturally to the  
mass distribution of compact objects similar to what is
currently observed. To eject the envelope, 
the NS rotational energy should be above
$\sim 10^{50}$ ergs (period of rotation 
$< 10$ ms). Unless
the magnetic coupling between the pre-collapse
core and envelope is strong enough to preclude
rapid rotation of the core, as was
suggested in Spruit and Phinney (1998), 
the magneto-rotational supernova explosion
is very attractive mechanism.    
This hypothesis has a number of addtional 
predictions, which can be verified by observations:

(1) Accretion-induced BH with masses of about 
$\Mov$ should exist. They could be detected in 
low-mass transient X-ray binaries.

(2) Supernova remnants containing BH must be less energetic 
evidencing less energetic supernova explosions.
Supernova remnants with NS should be axially symmetric.

(3) The coaxiality of the angular momentum and the
space velocity of a pulsar, as observed in Crab and Vela
pulsars, must be a common property of all radiopulsars.

(4) BH should have very small space velocities in comparison with 
pulsars.

Reliable measurements of the NS mass substantially above
1.6~M$_\odot$ would be 
\emph{a direct refutation of the proposed hypothesis}. 
\\[2mm]

%
The work is partially supported by RFBR grants 99-02-16205, 00-02-17164,
and 01-15-99310.
\\[3mm]


\indent
{\bf References\\[2mm]}
Ardeljan N.V., Bisnovatyi-Kogan G.S., Moiseenko S.G.: 2000, \AA \yr2000 \v355 \p1181 \\
Bisnovatyi-Kogan G.S.: 1970, \AZh \yr1970 \t47 \s813 \\
Blinnikov S.I., Novikov I.D., Perevodchikova T.V., Polnsrev A.G.: 1984, \PAZh \yr1984 \t10 \s177 \\
Brown G.E., Weingartner J.C., Wijers R.A.M.J.: 1996, \ApJ \yr1996 \v463 \p297 \\
Cherepashchuk A.M.: 1996, \UFN \yr1996 \t39 \s753 \\
Cherepashchuk A.M.: 1998, In Proc. Int. Conf. in Honour of Prof. A.G.Massevitch, 
"Modern Porblems of Stellar
Evolution", Ed. D.S.Wiebe, Moscow, \yr1998 \p198 \\
Cherepashchuk A.M.: 2000, {\it Space Sci. Rev.}, \yr2000 \v93 \p473 \\
Cherepashchuk A.M.: 2001, Astron. Rep. \yr2001 \v45 \p120 \\
Heap S.R., Corcoran M.F.: 1992, \ApJ \yr1992 \v387 \p340 \\
Herant M., Benz W., Hix J., Fryer C.L., Colgate S.A.: 1994, \ApJ \yr1994 \v435 \p339 \\
Imshennik V.S.: 1992, \PAZh \yr1992 \t18 \s489 \\
Janka H.-Th.: 2001, \AA \yr2001 \v368 \p527 \\
Mezzacappa A. et al.: 1998, \ApJ \yr1998 \v495 \p911 \\
Nice D.J., Splaver E.M., Stairs I.H.: 2001, \ApJ \yr2001 \v549 \p516 \\
Orosz J.A., Kuulkers E.: 1999, \MNRAS \yr1999 \v305 \p132 \\
Prakash M., Lattimer J.M., Pons J.A., Steiner A.W., Reddy S.: 2000, {\it astro-ph/0012136}. \\
Spruit H., Phinney S.: 1998, \Nat \yr1998 \v393 \p139
Stickland D., Lloyd C., Radzuin-Woodham A.: 1997, \MNRAS \yr1997 \v296 \p{L21} \\
Strobel K., Weigel M.K.: 2001, \AA \yr2001 in press (astro-ph/0012321). \\
Thorsett S., Chacrabarty D.: 1999, \ApJ \yr1999 \v512 \p288 \\
Timmes F.X., Woosley S.E., Weaver T.A.: 1996, \ApJ \yr1996 \v457 \p834 \\
van Kerkwijk M.H., van Paradijs J., Zuiderwijk E.J.: 1995, \AA \yr1995 \v303 \p497 \\
   
\end{document}